\documentstyle[preprint,aps,prb]{revtex}
\begin{document}
\draft \title{Soft X-ray Fluorescence Study
of Buried Silicides in Antiferromagnetically Coupled Fe/Si Multilayers}
\author{J.A. Carlisle, A. Chaiken, R.P. Michel and L.J. Terminello}
\address{Materials Science and Technology
Division\\Lawrence Livermore National Lab\\Livermore, CA 94551}
\author{J.J. Jia and  T.A. Callcott}
\address{University of Tennessee\\Knoxville, TN 37996}
\author{D.L. Ederer}
\address{Tulane University\\New Orleans, LA  70118}
\date{\today}

\maketitle

\begin{abstract}
Soft x-ray fluorescence spectroscopy has been employed to obtain
information about the Si-derived valence band states of Fe/Si
multilayers.  The valence band spectra are quite different for films
with and without antiferromagnetic interlayer exchange coupling,
demonstrating that these multilayers have different silicide phases in
their spacer layers.  Comparison with previously published
fluorescence data on bulk iron silicides shows that the Fe
concentration in the silicide spacer layers is substantial.  Near-edge
x-ray absorption data on antiferromagnetically coupled multilayers in
combination with the fluorescence data demonstrate unambiguously that
the silicide spacer layer in these films is metallic.  These results
on the electronic structure of buried layers in a multilayer film
exemplify the wide range of experiments made possible by new
high-brightness synchrotron sources.

\end{abstract}

\clearpage

\section{Introduction}

Multilayer films made by alternate deposition of two materials play an
important role in electronic and optical devices such as quantum-well
lasers and x-ray mirrors.\cite{chang} In addition, novel phenomena
like giant magnetoresistance and dimensional crossover in
superconductors have emerged from studies of multilayers.  While
sophisticated x-ray techniques are widely used to study the morphology
of multilayer films, progress in studying the electronic structure has
been slower.  The short mean-free path of low-energy electrons
severely limits the usefulness of photoemission and related electron
spectroscopies for multilayer studies.

Soft x-ray fluorescence (SXF) is a bulk-sensitive photon-in,
photon-out method to study valence band electronic
states.\cite{ederer} Near-edge x-ray absorption fine-structure
spectroscopy (NEXAFS) measured with partial photon yield can give
complementary bulk-sensitive information about unoccupied
states.\cite{stohrbook} Both these methods are element-specific since
the incident x-ray photons excite electrons from core levels.  By
combining NEXAFS and SXF measurements on buried layers in multilayers
and comparing these spectra to data on appropriate reference
compounds, it is possible to obtain a detailed picture of the
electronic structure.

The Fe/Si multilayer system well illustrates the power of combining
the SXF and NEXAFS techniques.  Fe/Si multilayers exhibit a large
antiferromagnetic (AF) interlayer exchange coupling that is apparently
similar to that previously observed in metal/metal multilayers like
Fe/Cr.\cite{fullerton} The observation of strong antiferromagnetic
coupling was initially surprising, since this coupling is believed to
be a manifestation of spin-density oscillations in the non-magnetic
metallic spacer layer of a multilayer.\cite{hathaway} The
interpretation of the Fe/Si coupling data was hampered by lack of
knowledge about the strongly intermixed iron silicide spacer layer,
which was variously hypothesized to be a metallic compound in the B2
CsCl structure\cite{fullerton} or a Kondo insulator in the more
complex B20 structure.\cite{mattson} If the spacer layer is not
metallic, then the usual theories of interlayer exchange coupling do
not apply\cite{hathaway} and the coupling must involve a novel
mechanism.  Using transmission electron microscopy (TEM), the spacer
layer has been identified as a metastable cubic iron silicide closely
lattice-matched to bulk Fe.\cite{fesiprb} However, since the exact
stoichiometry of the silicide was not determinable by diffraction
means, the question of whether the spacer layer is a metal or not has
remained unanswerable.  SXF and NEXAFS are ideal techniques to resolve
exactly this type of issue.

SXF and NEXAFS measurements were performed on five different Fe/Si
multilayer films at the Advanced Light Source on beamline 8.0, which
is described in detail elsewhere.\cite{beamline} SXF data has
previously been used to study buried layers of BN\cite{carlisle} and
Si.\cite{nilsson} Data taken at the Fe L-edge closely resembles bulk
Fe for all Fe/Si multilayers.  NEXAFS spectra were acquired by
measuring the total Si L-emission yield with the same detector used
for fluorescence.  The resulting data are expected to be comparable to
those acquired by electron counting.\cite{stohrbook} The films used in
this study were grown using ion-beam sputtering (IBS) in a chamber
with a base pressure of 2$\times$10$^{-8}$ torr.  The deposition
conditions were the same as those used in previous
studies.\cite{fesiprb} All multilayers were characterized using x-ray
diffraction and magnetometry.  Reference spectra were obtained from a
crystalline silicon (c-Si) substrate piece, an amorphous silicon
(a-Si) film made by IBS, and a fragment of an $\rm FeSi_2$ sputter
target.


\section{Results}

Figure~\ref{magcurves} shows hysteresis loops for three representative
Fe/Si multilayers.  The polycrystalline (Fe30\AA/Si20\AA)x50
multilayer grown on glass has a magnetization curve that shows no sign
of interlayer exchange coupling.  This multilayer has magnetic
properties like those of bulk Fe.  The epitaxial (Fe40\AA/Si14\AA)x40
multilayer grown on MgO has a low remanent magnetization and a high
saturation field, which are the classic signs of antiferromagnetic
interlayer coupling.  Data on the polycrystalline (Fe30\AA/Si14\AA)x50
multilayer fall somewhere in-between these two extremes.  Detailed
characterization of these films has been published
previously.\cite{fesiprb}

For purposes of comparison to the Fe/Si multilayer SXF spectra, SXF
reference spectra taken at the Si L-edge for the c-Si and a-Si samples
are shown in Fig.~\ref{refspect}.  The spectra resemble previously
published Si data.\cite{ederer,nilsson} The peaks near 89 and 92 eV in
the c-Si spectrum originate from non-bonding $s$ states and
$sp$-hybridized states, respectively.\cite{nilsson,jia} These features
are broadened by disorder in a-Si.

Figure~\ref{allxrf} shows the Si L-edge valence band emission spectra
of the $\rm FeSi_2$ reference sample and the same two polycrystalline
Fe/Si multilayers whose magnetization data are shown in
Figure~\ref{magcurves}.  The $\rm FeSi_2$ data has two primary
features, namely $s$-orbital features near 90 eV, and a shoulder which
extends up to 99 eV and is comprised mostly of states with $d$
symmetry.  These features have been previously identified in
semiconducting bulk $\rm FeSi_2$ specimens.\cite{jia}

In Fig.~\ref{allxrf} the spectrum for the polycrystalline
antiferromagnetically coupled multilayer with $\rm t_{Si}$ = 14\AA\
looks similar to the $\rm FeSi_2$ data, while the spectrum for the
polycrystalline uncoupled multilayer with $\rm t_{Si}$ = 20\AA\ is
more like c-Si.  Peaks in the AF-coupled multilayer spectrum are
noticeably narrower than those in the $\rm FeSi_2$ reference spectrum.
Studies of bulk iron silicides have shown that peaks in the Si
emission spectra narrow as the iron content increases and Si-Si
coordination decreases.\cite{jia} Thus the data of Fig.~\ref{allxrf}
indicate that the Fe atomic fraction in the spacer layer of the
AF-coupled multilayers is higher than 1/3.  Overall the shape of the
spectrum from the AF-coupled multilayer is more reminiscent of SXF
data on bulk B20 FeSi than of data on bulk $\rm FeSi_2$.\cite{jia} The
uncoupled multilayer data in Fig.~\ref{allxrf} have a sharp peak near
92 eV which coincides with a feature in the c-Si spectrum although the
shape of the higher energy part of the valence band more closely
resembles the $\rm FeSi_2$ data.  The narrowness of the 92 eV feature
is evidence for a significant Fe content and low Si-Si coordination in
the spacer layer of the uncoupled multilayer.  These observations are
consistent with the TEM determination that the spacer layer in the
uncoupled multilayers is amorphous iron silicide.\cite{fesiprb}


The presence of significant Fe in the silicide spacer layer of the
Fe/Si multilayers strongly suggests that the silicide is metallic.
Unambiguous confirmation of the metallic nature of the silicide is
obtained by plotting together the SXF and NEXAFS spectra as in
Fig.~\ref{nexafs}.  For this data set the spectrometer energy
calibration was accomplished through comparison with earlier work on
c-Si L-emission\cite{rubensson} and through alignment of the
elastically scattered photon peak to the incident photon energy.  The
overlap of the valence band features from the SXF and the conduction
band features from NEXAFS is therefore convincing evidence that the
silicide spacer layer of the multilayer is metallic.  While the Si
bands near the Fermi level clearly show the energy gap which is
expected in a semiconductor, the slope of the silicide bands near
$E_F$ suggests that the Fermi level falls in the middle of an energy
band.  A more detailed interpretation of these spectral features will
require electronic structure calculations.

\section{Discussion and Conclusions}

SXF data have also been taken on an Fe/Si multilayer with $\rm t_{Si}$ =
14\AA\ but which was held at a reduced temperature of 120 K during
growth (data not shown).  The valence band spectra of the film grown
at reduced temperature with $\rm t_{Si}$ = 14\AA\ look virtually identical
to data on the film grown at 60$^{\circ}$C but with $\rm t_{Si}$ = 20\AA.
The most likely explanation for this similarity is that both films
have amorphous iron silicide spacer layers.  The amorphous state of
the spacer layer in these films must be due to the reduced Fe content
compared with films which have thinner Si layers or are deposited at
higher temperature.  Multilayers with amorphous spacer layers do not
display antiferromagnetic interlayer coupling.\cite{fullerton,fesiprb}

A comparison of the data of Figs.~\ref{allxrf} and \ref{nexafs} show
that the peaks in the spectrum of the epitaxial AF-coupled multilayer
are narrower than those in the spectrum of the polycrystalline
AF-coupled multilayer.  This suggests that a higher degree of local
order occurs in epitaxial films.  The nature of this order and the
exact structure of the silicide spacer layer phase are not yet known.
TEM studies have shown that the spacer layer in AF-coupled multilayers
is a crystalline cubic iron silicide in the B2 CsCl phase or fcc $\rm
DO_3$ phase.\cite{fesiprb} The TEM diffraction patterns are not
consistent with the B20 structure, whose SXF data most closely
resembles that of the AF-coupled multilayers.  Jia {\it et al.} do
report SXF data on the $\rm DO_3$-structure $\rm Fe_3Si$ phase but the
spectrum of this compound has a much more prominent and narrow
non-bonding $s$ feature.\cite{jia} The presence of an $\rm Fe_3Si$
spacer can be ruled out on other grounds since this compound is
ferromagnetic, inconsistent with the presence of antiferromagnetic
interlayer coupling.  The possibility remains, however, that the
spacer layer is in the $\rm DO_3$ structure but at a different
stoichiometry.  No SXF data on the metastable B2 silicide phase have
been reported although photoemission measurements show that it is
metallic.\cite{vonkanel} The magnetic properties of the B2 phase and
hypothetical off-stoichiometry phases are not known.  The observation
of large biquadratic coupling in Fe/Si
multilayers\cite{fullerton2,michel} suggests that an antiferromagnetic
or ferrimagnetic order may be present in the spacer layer.

When examined together, the SXF and NEXAFS data show that Fe/Si
multilayers with crystalline metallic silicide spacer layers have
antiferromagnetic interlayer coupling, while similar multilayers with
amorphous silicide spacer layers show no interlayer coupling.  Whether
the amorphous silicide layers are metallic or semiconducting is a
topic for further study.  Theoretical calculations will be necessary
to get a better estimate of the stoichiometry and magnetic properties
of the silicide spacer in the AF-coupled multilayers.  The present
data should lay to rest any speculation that the interlayer exchange
coupling in Fe/Si multilayers involves a novel mechanism.  The clarity
of these results on thin buried silicide layers illustrates the power
of photon-counting spectroscopies with their intrinsic bulk
sensitivity for the study of multilayer films.

\bigskip

\paragraph*{Acknowledgements}

\noindent We would like to thank P.E.A. Turchi, P.A. Sterne and J. van
Ek for helpful discussions.  This work was supported by the Division
of Materials Science, Office of Basic Energy Sciences, and performed
under the auspices of the U.S. Department of Energy by Lawrence
Livermore National Laboratory under contract No.  W-7405-ENG-48, by
National Science Foundation Grant No. DMR-9017996 and DMR-9017997, by
a Science Alliance Center for Excellence Grant from the University of
Tennessee, by the U.S. Department of Energy (DOE) Contract No. DE-
AC05-84OR21400 with Oak Ridge National Laboratory and by the Louisiana
Educational Quality Support Fund and DOE-EPSCOR Grant LEQSF (93-95)-03
at Tulane University.  This work was performed at the Advanced Light
Source, which is also supported by the Office of Basic Energy
Sciences, U.S Department of Energy, under contract
No. DE-AC03-76SF00098.

\clearpage


\clearpage

\begin{figure}
\caption{Magnetization curves for three Fe/Si
multilayers.  The y-axis shows magnetization data normalized to the
saturated value.  The solid line indicates data for a polycrystalline
(Fe30\AA/Si20\AA)x50 multilayer which has a magnetization curve much
like bulk Fe.  The open circles indicate data for an epitaxial
(Fe40\AA/Si14\AA)x40 multilayer which has the high saturation field
and low remanent magnetization that are characteristic of
antiferromagnetic interlayer exchange coupling.  The polycrystalline
(Fe30\AA/Si14\AA)x50 multilayer (indicated by filled circles) has
weaker antiferromagnetic coupling than the epitaxial multilayer.}
\label{magcurves}
\end{figure}

\begin{figure}
\caption{SXF Si L-emission spectra for crystalline and amorphous silicon
films.  These data were taken with an incident photon energy of 132
eV.}
\label{refspect}
\end{figure}

\begin{figure}
\caption{SXF Si L-emission spectra for an $\rm FeSi_2$ reference
sample and for the two polycrystalline Fe/Si multilayers whose
magnetization curves are shown in Figure 1.  The incident photon
energy was 132 eV.  The data labelled ``Uncoupled ML'' is from the
(Fe30\AA/Si20\AA)x50 multilayer grown on glass.  The data labelled
``AF-coupled ML'' is from the antiferromagnetically coupled
(Fe30\AA/Si14\AA)x50 multilayer grown on glass.}
\label{allxrf}
\end{figure}


\begin{figure}
\caption{SXF Si L-emission spectra (solid line) and Si L-edge NEXAFS
(dashed line) for the crystalline Si reference film and for the
epitaxial (Fe40\AA/Si14\AA)x40 multilayer on MgO.  The crossing of the
valence band data obtained from SXF and the conduction band data
obtained from NEXAFS demonstrates that the silicide spacer layer is
metallic.}
\label{nexafs}
\end{figure}

\end{document}